# Graphene-Based Electromechanical Thermal Switches

Michelle E. Chen[1], Miguel Muñoz Rojo[2,3], Feifei Lian[3], Justin Koeln[4,5], Aditya Sood[1], Stephanie M. Bohaichuk[3], Christopher M. Neumann[3], Sarah G. Garrow[4], Andrew G. Alleyne[4], Kenneth E. Goodson[1,6], Eric Pop[1,3,*]

[1]*Dept. of Materials Science & Eng., Stanford University, Stanford, CA 94305, U.S.A.*
[2]*Dept. of Thermal & Fluid Eng., University of Twente, Enschede, 7500 AE, Netherlands.*
[3]*Dept. of Electrical Eng., Stanford University, Stanford, CA 94305, U.S.A.*
[4]*Dept. of Mechanical Sci. & Eng., University of Illinois Urbana-Champaign, Urbana, IL 61801, U.S.A.*
[5]*Dept. of Mechanical Eng., University of Texas at Dallas, Richardson, TX 75080, U.S.A.*
[6]*Dept. of Mechanical Eng., Stanford University, Stanford, CA 94305, U.S.A.*

Thermal management is an important challenge in modern electronics, avionics, automotive, and energy storage systems. While passive thermal solutions (like heat sinks or heat spreaders) are often used, actively modulating heat flow (e.g. via thermal switches or diodes) would offer additional degrees of control over the management of thermal transients and system reliability. Here we report the first thermal switch based on a flexible, collapsible graphene membrane, with low operating voltage, < 2 V. We also employ active-mode scanning thermal microscopy (SThM) to measure the device behavior and switching in real time. A compact analytical thermal model is developed for the general case of a thermal switch based on a double-clamped suspended membrane, highlighting the thermal and electrical design challenges. System-level modeling demonstrates the thermal trade-offs between modulating temperature swing and average temperature as a function of switching ratio. These graphene-based thermal switches present new opportunities for active control of fast (even nanosecond) thermal transients in densely integrated systems.

**1. Introduction**

Advances in modern technology have been accompanied by a surge in energy consumption and a growing need to control energy dissipation of electronics, from mobile devices to data centers [1]. Controlling energy lost as waste heat is not only desirable for increasing the energy efficiency of electronics but also critical for improving device reliability and lifetime [2]. Modern thermal management methods for electronics often include macroscale heat exchangers, such as heat sinks, heat pipes, or phase change approaches [3, 4]. These examples can be understood as *passive* thermal components, similar to thermal resistors and thermal capacitors. However, when compared to analogous electrical devices, thermal management is limited by a lack of *active* thermal devices, such as thermal transistors, switches, or diodes [5], that would be capable of manipulating heat flow in a controlled manner, similar to the routing of electricity.

A fundamental difference between active electrical components and (the relative lack of) active thermal components is that electrons obey Fermi-Dirac statistics, meaning their Fermi level can be manipulated by a gating voltage. However, heat in electronics is typically carried by lattice vibrations (phonons) which obey Bose-Einstein statistics, and cannot be "gated." Instead, phonons could be manipulated by differences in temperature, density of states, mass density [6] or by geometrical and mechanical methods such as spatial confinement and physical switching.

Among active thermal devices, thermal switches could offer the ability to regulate temperature transients and reduce thermal fatigue over concentrated regions. A thermal switch relies on non-thermal parameters such as electric field, electrochemical potential, or pressure, to alter the device thermal conductance [5]. Several technologies have been reported for thermal switching, including liquid metal actuation [7, 8], ion intercalation between layered materials [9], externally biased phase change materials [10], and Micro-Electro-Mechanical Systems (MEMS) [11-14]. However, these devices have typically low thermal switching ratios or slow operation, which limits their potential use.

In this work, we demonstrate the first active thermal switches based on reversible, collapsible graphene membranes. These novel devices operate at low voltage (< 2 V) and could be reduced to nanoscale dimensions, operating at lower power and higher frequency. In comparison, similar thermal switches have been made with electrostatically collapsible metal membranes [11-14], however their utility is limited by high operating voltages, from 12 V to 126 V, in part due to the thickness of the metal membranes used. In contrast, graphene is an electrically and thermally conductive two-dimensional (2D) layer of carbon atoms that is ~3.35 Å "thick," with the highest intrinsic tensile strength, stiffness, and in-plane thermal conductivity (2000 to 4000 Wm$^{-1}$K$^{-1}$ when suspended) of any material, comparable only to that of carbon nanotubes and diamond [15, 16]. Graphene has already been demonstrated as a promising material in nanoscale electro-mechanical switches (NEMS) [17-19], yet despite its high thermal conductivity, it has not been previously explored as a thermally conductive switching membrane.

## 2. Experimental Work

Figure 1(a) shows an illustrated schematic of the graphene thermal switch device. Graphene micro-ribbons are suspended over thermally and electrically insulating legs between top (metal) and bottom (silicon) electrodes. In this 'off' state, the device demonstrates limited heat flow in the cross-plane direction. The switch is turned 'on' by applying a bias between the top and bottom electrodes to electrostatically deflect the graphene until it contacts the underlying silicon electrode. In the 'on' state, the locally suspended graphene membranes become channels for additional cross-plane heat flow. When the electrical bias is removed, the elastic restoring force of graphene causes the membrane to suspend, returning the device to the off state.

### 2.1 Device Fabrication

Our devices were fabricated using high-quality monolayer graphene grown by chemical vapor deposition (CVD) [20-22]. We sequentially transferred two layers (2L) of CVD graphene onto 540 nm thick thermally grown SiO$_2$ on highly doped (*n*-type, 1 to 5 mΩ·cm) Si substrates (Supplementary section 1). The active region of the graphene devices were defined using optical photolithography, a copper hard mask, and O$_2$ plasma etching leaving graphene channels free of photoresist residue. Top electrodes consisting of a 3 nm Cr sticking layer and 40 nm Au were deposited by electron beam evaporation which clamped the graphene to the substrate. This served both as an etch mask for the $SiO_2$ and as a top electrode for the final device. Approximately 500 or 540 nm of unmasked $SiO_2$ was removed with 20:1 buffered oxide etch (Supplementary section 2), releasing the graphene membranes, which were then dried immediately using a critical point dryer. The resulting graphene structures were thus suspended over $SiO_2$ legs without collapse. An additional type of device was fabricated in which 3 nm thick Cr lines in varying geometries were deposited over the graphene.

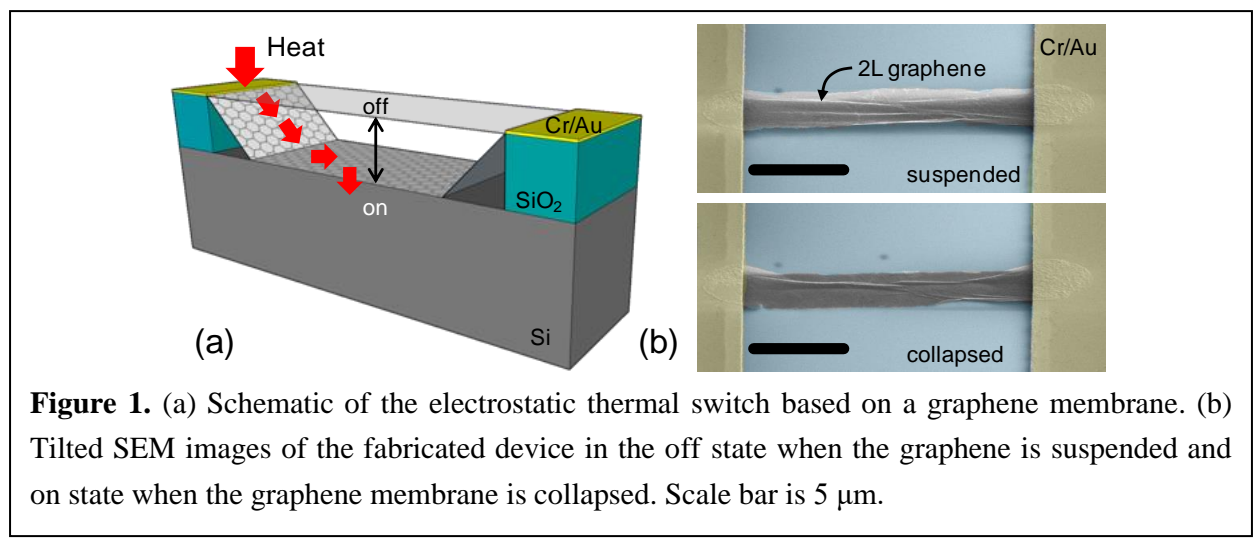

**Figure 1.** (a) Schematic of the electrostatic thermal switch based on a graphene membrane. (b) Tilted SEM images of the fabricated device in the off state when the graphene is suspended and on state when the graphene membrane is collapsed. Scale bar is 5 μm.

For electrical characterization, the underlying oxide was fully removed so that the graphene could electrically contact the underlying highly doped silicon, which served as a bottom electrode. For devices characterized thermally, the remaining 40 nm of $SiO_2$ were left to electrically insulate the SThM probe and circuit from the in-situ electrical measurement setup (Supplementary section 2). The graphene devices range in length from 10 to 24 μm, and their suspension was verified using tilted scanning electron microscopy (SEM), as shown in Figure 1(b). Suspension and collapse of the electrically actuated membrane was also observed under an optical microscope during switching.

## 2.2. Thermal Measurements

The design and dimensions of our graphene-based devices present a unique thermal metrology challenge. Due to structures that are <5 μm in lateral dimension, the spatial resolution of common optical characterization techniques such as infrared (IR) microscopy or time domain thermoreflectance (TDTR) is insufficient. Confocal Raman thermometry has sub-micron spatial resolution [23], however it cannot be applied to the metal regions at our device contacts, which is necessary to thermally characterize the heat flow in both the off (suspended) and on (collapsed) states. We therefore use scanning thermal microscopy (SThM), with ~100 nm spatial resolution, to evaluate heat flow in such NEMS devices for the first time.

SThM is an atomic force microscope (AFM) technique that uses a V-shaped tip whose electrical resistance is a function of temperature. This probe tip can act simultaneously as a heater for our measured device, and as a temperature-dependent variable resistor within a Wheatstone bridge circuit [24-27]. When the device is switched, the increase of cross-plane heat flow through the collapsed graphene causes a corresponding temperature decrease and electrical resistance change of the SThM probe tip, which is reflected in the change of SThM circuit voltage, $\Delta V_{SThM}$. (Additional details of SThM are provided in Supplementary section 3.)

As shown in the schematic diagram in Figure 2(a), we place the SThM probe tip in contact with our top electrode at a fixed location near the graphene-metal junction. We employ active-mode

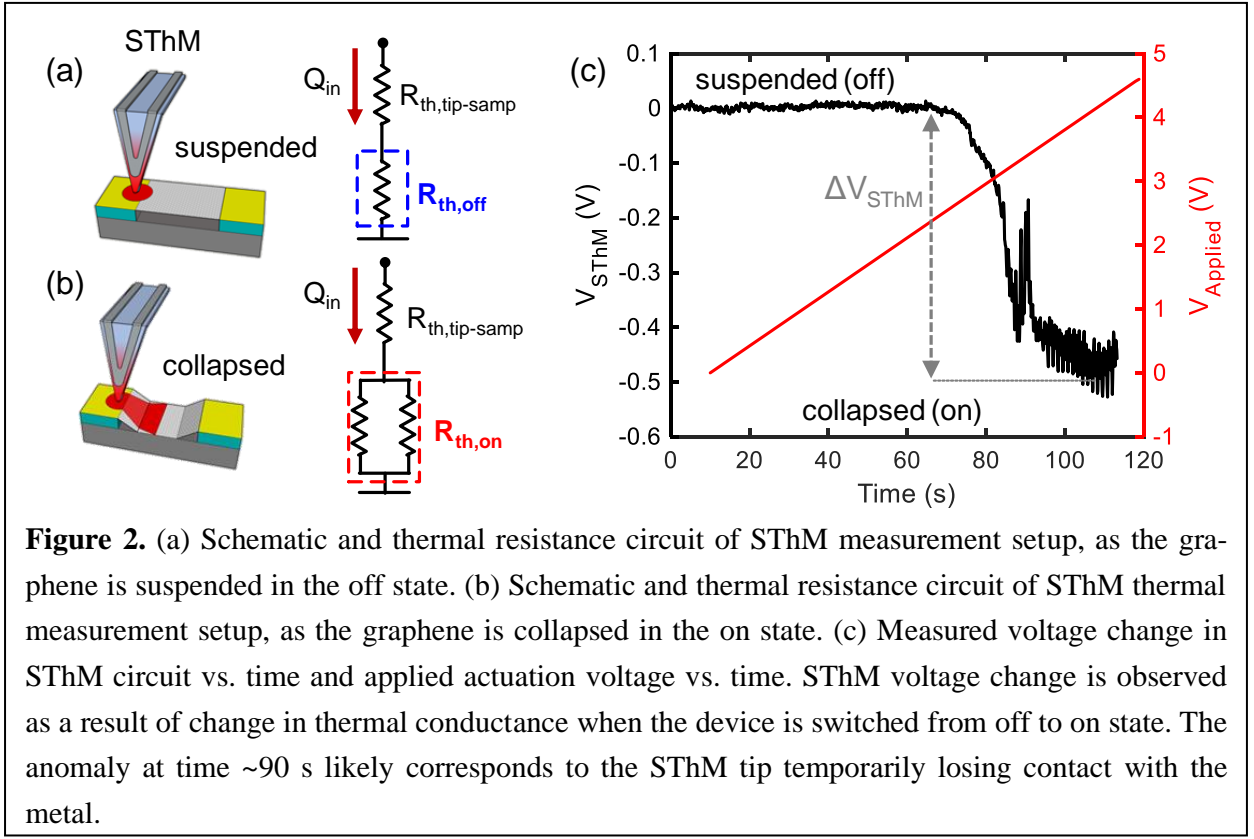

**Figure 2.** (a) Schematic and thermal resistance circuit of SThM measurement setup, as the graphene is suspended in the off state. (b) Schematic and thermal resistance circuit of SThM thermal measurement setup, as the graphene is collapsed in the on state. (c) Measured voltage change in SThM circuit vs. time and applied actuation voltage vs. time. SThM voltage change is observed as a result of change in thermal conductance when the device is switched from off to on state. The anomaly at time ~90 s likely corresponds to the SThM tip temporarily losing contact with the metal.

SThM whereby the tip is electrically heated at a constant power, and we wait for the tip to reach thermal steady state. When we ramp the voltage up to 5 V between the top and bottom electrodes of our thermal switch, the graphene membrane collapses as depicted in Figure 2(b), inducing a measurable voltage change of the SThM tip, as shown in Figure 2(c). This represents clear evidence of the dynamically changing heat flow path from the SThM tip, through the graphene membrane, and into the substrate. Figures 2(a) and 2(b) display a schematic of the thermal circuit in the off (suspended membrane) and on states (collapsed membrane). Because we are measuring *differences* in tip voltage, $\Delta V_{SThM}$, these measurements automatically eliminate extrinsic thermal effects (such as thermal convection, thermal radiation, and thermal contact resistance) between the off and on states of the device. (Additional details in Supplementary section 3.)

## 3. Results and Discussion

From the SThM voltage changes between the off and on states of the graphene membrane device, we calculate the resultant thermal switching ratios as follows:

$$\Delta T = Q_{in} \mathcal{R}_{th} = \frac{1}{\alpha}\left(\frac{R_{probe}}{R_0} - 1\right) \tag{1}$$

where $\Delta T$ is the temperature rise of the SThM tip above ambient, $Q_{in}$ is the electrical power heating the SThM tip, $\mathcal{R}_{th}$ is the thermal resistance between SThM tip and thermal ground [Figure 2(c)], $R_0$ is the electrical resistance of the SThM probe at room temperature, and $\alpha$ is the temperature coefficient of resistance of the Pd tip [28]. Then, the off/on thermal switching ratio ($\mathcal{R}_{ratio}$) of the device is

$$\mathcal{R}_{ratio} = \frac{\mathcal{R}_{th,off}}{\mathcal{R}_{th,on}} = \frac{\Delta T_{off}}{\Delta T_{on}} = \frac{\frac{R_{probe,off}}{R_0}-1}{\frac{R_{probe,on}}{R_0}-1} \qquad (2)$$

where $R_{probe}$ is the measured electrical resistance of the SThM tip.

We measured 27 devices made with a graphene switching layer and found that the mean thermal switching ratio for such devices was 1.08 ± 0.02, and the length of the device showed no impact on thermal performance of the switch. While longer devices were anticipated to have favorable low switching voltages, the thermal impact of the collapsed graphene region in the on state is most likely limited by the thermal healing length of graphene on silicon, which has been calculated to be approximately 0.1 to 0.2 μm for graphene on $SiO_2$ [16].

We placed the SThM probe tip within 1 μm accuracy at the edge of our top electrode immediately adjacent to the graphene switching membrane. We carried out finite element simulations to determine the effect of SThM probe placement and collapse length of the switching membrane on the thermal measurement (Supplementary section 3). Our models considered the cases of the tip placement immediately at the center of the 5 μm wide top electrode (2.5 μm away from the graphene), and near the edge of the graphene-electrode junction (200 nm away from the graphene, corresponding to the diameter of the SThM tip). Given that during measurements the SThM tip was placed within 1 μm of the graphene-electrode junction, it is estimated from the finite element model (Supplementary section 3) that the error from tip placement in measuring the thermal switching ratio is <0.02.

### 3.1. Compact Analytical Model

To gain physical insight for optimizing the thermal switch design, we developed a compact analytical model of the thermal switching ratio. The thermal switching ratio is defined as the ratio of the off-state thermal resistance to the on-state thermal resistance, given by:

$$\mathcal{R}_{ratio} = \frac{\mathcal{R}_{th,off}}{\mathcal{R}_{th,on}} \approx 1 + \frac{t_{stack}}{k_{stack} \cdot L_{contact}} \left( \frac{\mathcal{R}'_{th,boundary}}{L_{collapse}} + \frac{L_{total} - L_{collapse}}{k_{membrane} t_{membrane}} \right)^{-1} \qquad (3)$$

where $t_{stack}$ is the height of the insulating pillar and $k_{stack}$ its thermal conductivity, while $t_{membrane}$ and $k_{membrane}$ refer to the thickness and thermal conductivity of the switching membrane, respectively, illustrated in Figure 3(a). $L_{contact}$ is the length of the metal contact clamping down the switching membrane and $L_{total}$ is the total length of the free-standing membrane in the off state. $L_{collapse}$ is the length of the collapsed region after the device is switched on, and is a function of the applied voltage based on an electrostatic model (Supplementary section 4). The thermal boundary resistance of the collapsed graphene with $SiO_2$ is $\mathcal{R}_{th,boundary} \approx 2\times10^{-8}$ m$^2$KW$^{-1}$ [6, 29].

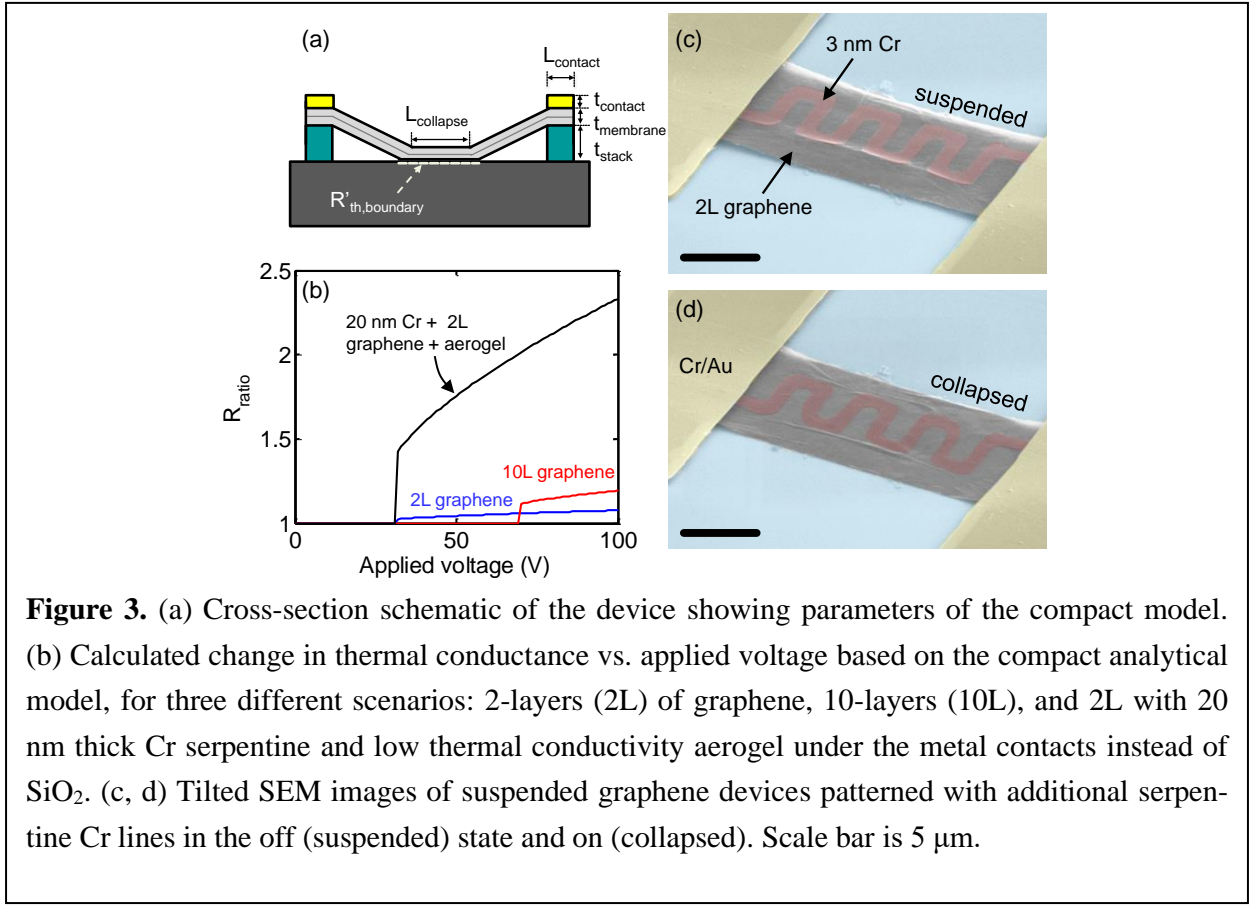

**Figure 3.** (a) Cross-section schematic of the device showing parameters of the compact model. (b) Calculated change in thermal conductance vs. applied voltage based on the compact analytical model, for three different scenarios: 2-layers (2L) of graphene, 10-layers (10L), and 2L with 20 nm thick Cr serpentine and low thermal conductivity aerogel under the metal contacts instead of $SiO_2$. (c, d) Tilted SEM images of suspended graphene devices patterned with additional serpentine Cr lines in the off (suspended) state and on (collapsed). Scale bar is 5 μm.

We used our thermal model in conjunction with a modified electrostatic model developed by Bao *et al.* [30] (Supplementary section 4) to simulate the expected thermal switching ratios as a function of applied voltage. Figure 3(b) shows the abrupt change in thermal switching ratio at the mechanical pull-in condition. We note that while the measured thermal switching ratio is comparable to the results of the compact model, the measured switching voltage is significantly lower than the predicted values. This observed low voltage switching is most likely due to a graphene membrane which was not initially taut, whereas the model is based on a taut membrane with no initial deflection. For example, during the polymer-assisted transfer process wrinkles are introduced to the graphene on a flat substrate. Upon suspension and release, the wrinkles unfold under the strain of the now doubly-clamped suspended graphene membrane [31].

We derived our device design principle based on the compact model, with consideration for electrical and thermal parameters. From an electromechanical perspective, we designed the suspended structure to have low height-to-width aspect ratio, and a thin (sub-nanometer) switching membrane in order to minimize the actuation voltage. However, from a thermal perspective, our model shows that it is critical to minimize off-state thermal leakage, either through materials selection or by increasing the device height, and a thicker switching membrane to maximize on-state thermal conductance.

Our compact model predicts that while a device with a membrane consisting of multiple (~5) layers of graphene has an improved thermal switching ratio, it is also predicted to have a significantly increased switching voltage compared to a device with only 2 graphene layers. To increase the switching membrane thickness and thermal conductance in a facile manner, we deposited additional lines of 3 nm thick chromium (Cr) over the graphene in various geometries (Supplementary section 5). We made devices with two linear Cr beams over the graphene, and were able to measure an improved thermal switching ratio of 1.10 (Supplementary Figure S6).

We also patterned serpentine Cr structures over the graphene [Figure 3(c, d)], which increased the coverage of metal on graphene while remaining flexible enough to not significantly increase the switching voltage of the device. These devices had an average thermal switching ratio of 1.22, with an actuation voltage of 3 V. The graphene integration in such a NEMS structure is critical, because the taut underlying graphene allows the flexible serpentine metal to remain suspended. Several conditions can be applied to this model for designing a switch with optimized thermal performance. For example, if $SiO_2$ is replaced with low thermal conductivity dielectrics such as porous silica or alumina aerogels with thermal conductivity of ~0.1 $Wm^{-1}K^{-1}$, and serpentine metal lines of 20 nm thick chromium are patterned over 50% of the graphene, a thermal switching ratio of over 2 could be achieved according to our calculations, as shown in Figure 3(b).

**3.2 System Level Simulations**

There are several temperature-dependent failure mechanisms in microelectronics, including packaging thermomechanical failure, metal diffusion, and leakage currents which exponentially increase with temperature [2]. One promising application of thermal switches is thermal regulation, as discussed in Ref. [5]. When deployed as a thermal regulator, a thermal switch is able to reduce the temperature fluctuation of a microelectronic component or system undergoing time-varying heat loads, thus reducing its thermal fatigue [5]. This section analyzes the thermal system and periodic operating condition shown in Figure 4(a) to provide insight into the impacts of the thermal switching ratio on temperature regulation. The system consists of a periodic heat source $Q$, which represents the heating of a CPU or power inverter, with period $P$ and duty-cycle $D$, where $0 < D < 1$. This heating source has a temperature $T_d$ and a thermal capacitance $C_d$. The thermal switch with thermal resistance, $\mathcal{R}_{th}$, is placed between the heat source and a heat sink with constant temperature, $T_s$. To achieve thermal regulation, the thermal switch is operated as shown in Figure 4(b), such that it is "on" during the heat load, $\mathcal{R}_{th} = \mathcal{R}_{th,on}$, and "off" in between heat loads, $\mathcal{R}_{th} = \mathcal{R}_{th,off}$. Thus, the thermal switch is capable of reducing the amplitude of the temperature oscillations, $T_{amp} = T_{max} - T_{min}$, when compared to the case where the thermal switch is always on. However, this reduction in amplitude comes at the cost of increasing the average temperature of the system, $T_{ave} = 0.5(T_{max} + T_{min})$. Thus, the goal is to design a thermal switch that minimizes $T_{amp}$ without significantly increasing $T_{ave}$.

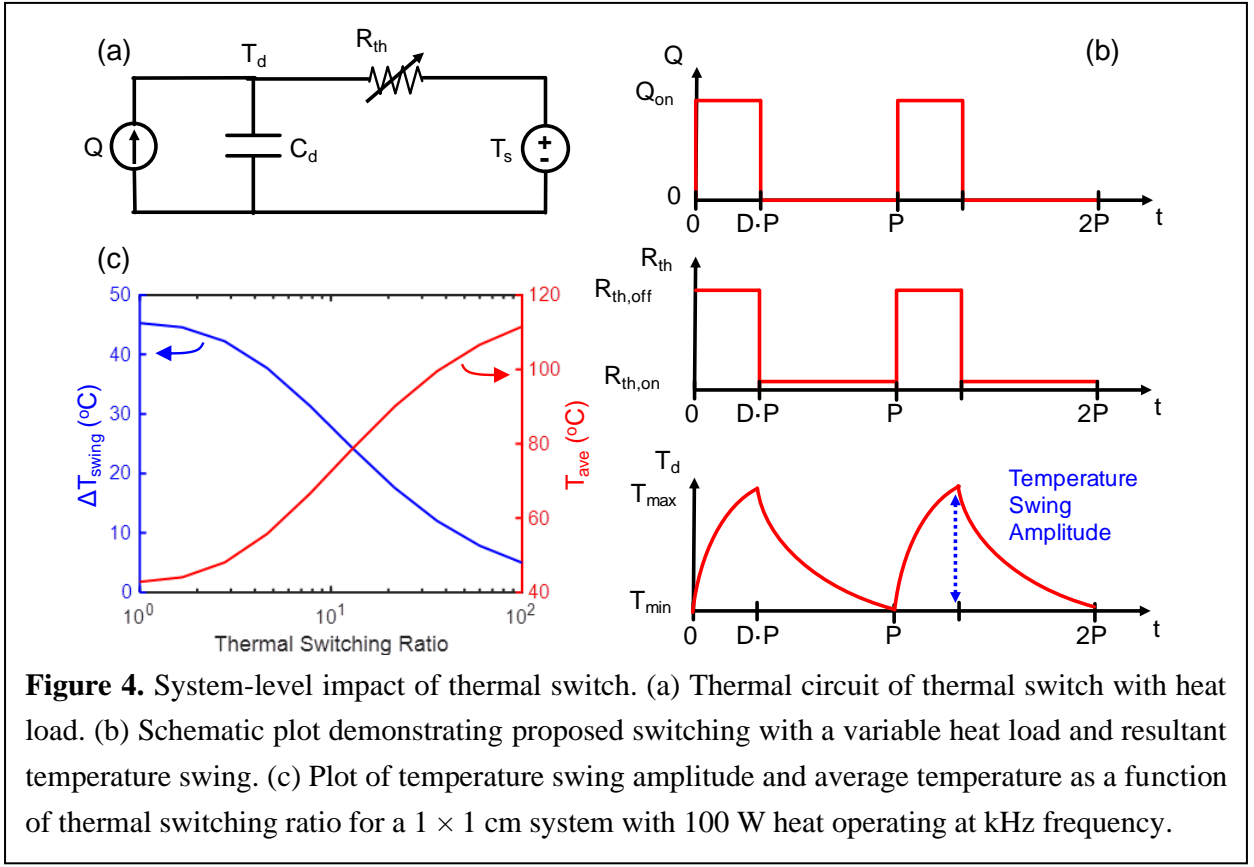

**Figure 4.** System-level impact of thermal switch. (a) Thermal circuit of thermal switch with heat load. (b) Schematic plot demonstrating proposed switching with a variable heat load and resultant temperature swing. (c) Plot of temperature swing amplitude and average temperature as a function of thermal switching ratio for a 1 × 1 cm system with 100 W heat operating at kHz frequency.

To determine the effects of the thermal switch ratio, $\mathcal{R}_{ratio}$, on $T_{amp}$ and $T_{ave}$, the differential equation for the device temperature is given by

$$C_\text{d} \frac{dT_\text{d}}{dt} = Q - \frac{1}{\mathcal{R}_\text{th}}(T_\text{d} - T_\text{s}). \tag{4}$$

Analyzing each of the two modes of operation it is possible to derive the following analytical relationships for $T_\text{amp}$ and $T_\text{ave}$:

$$T_\text{amp} = \mathcal{R}_\text{th,on} Q_\text{on} \frac{(1-e^{-\theta\varphi})(1-e^{-\theta})}{(1-e^{-\theta(1+\varphi)})} \tag{5}$$

$$T_\text{ave} = T_\text{s} + \frac{1}{2}\mathcal{R}_\text{th,on} Q_\text{on} \frac{(1+e^{-\theta\varphi})(1-e^{-\theta})}{(1-e^{-\theta(1+\varphi)})} \tag{6}$$

where $\theta = DP/(C_\text{d}\mathcal{R}_\text{th,on})$ and $\varphi = \mathcal{R}_\text{ratio}(1/D - 1)$. While these relationships are applicable to any thermal switch design and operation, it is valuable to consider a particular case to observe the benefit of the thermal switch. For a $1 \times 1$ cm area representative of the periodically heated system, the parameters can be approximated as $Q_\text{on} = 100$ W, $P = 10^{-3}$ s, $D = 0.1$, $C_\text{d} = 1.7\times 10^{-4}$ J/K, and $T_\text{s} = 20°$C. Assuming the thermal switch is designed to achieve on-state resistivity $\mathcal{R}'_\text{th,on} = 10^{-4}$ m$^2$KW$^{-1}$ (per area), Figure 4(c) shows the effect of $\mathcal{R}_\text{ratio}$ on $T_\text{amp}$ and $T_\text{ave}$.

The results of this model demonstrate there is an optimal switching ratio where the temperature swing amplitude can be reduced at the cost of increasing the average temperature. Figure 5c illustrates that with no thermal switching ($\mathcal{R}_\text{ratio} = 1$) thermal spikes in the form of temperature swing amplitude may exceed 40°C. As $\mathcal{R}_\text{ratio}$ increases, the temperature swing decreases while the average temperature increases. However if $\mathcal{R}_\text{ratio}$ is too high, the average temperature of the system can increase to the point of inducing failure (e.g. interconnect failure in microelectronics [2]). Under the above state conditions, in this model we find that $\mathcal{R}_\text{ratio} > 10$ would not be optimal from a thermal budget perspective. The parameters also highlight the significance of geometry, materials, and switching speed on the system-level impact of thermal switches (Supplementary section 6). The equivalent thermal resistance of the thermal switch is ideally low in the on- and high in the off-state for a better switching ratio. However, if either state has high thermal resistance, the average temperature of the system will rise. Moreover, at higher switching speeds the thermal switch becomes limited by its thermal time constant and the thermal switching ratio has less impact on the temperature swing amplitude. Both of these results highlight the need for high thermal conductivity materials with low thermal boundary resistance interfaces in the design and implementation of thermal switches.

### 3.3. Reversible Cycling

Reversible cycling of the thermal switch was verified by electrical measurements and thermal testing by SThM. Voltage pulses of 1.5 V amplitude and 30 s width were applied to the device and the cross-plane current through the graphene to the Si substrate was measured, as shown in Figure 5(a). (The cross-plane electrical measurement was enabled by extending the liquid etch to remove all underlying $SiO_2$, allowing the graphene to contact the underlying Si.) A gradual decrease of on-state current is attributed to electrostatic attraction of ambient particles and possible damage of the graphene at the edge contacts. However, we can verify the graphene is fully suspended in the off-state, as the off-state current reaches the noise floor of the measurement (~pA).

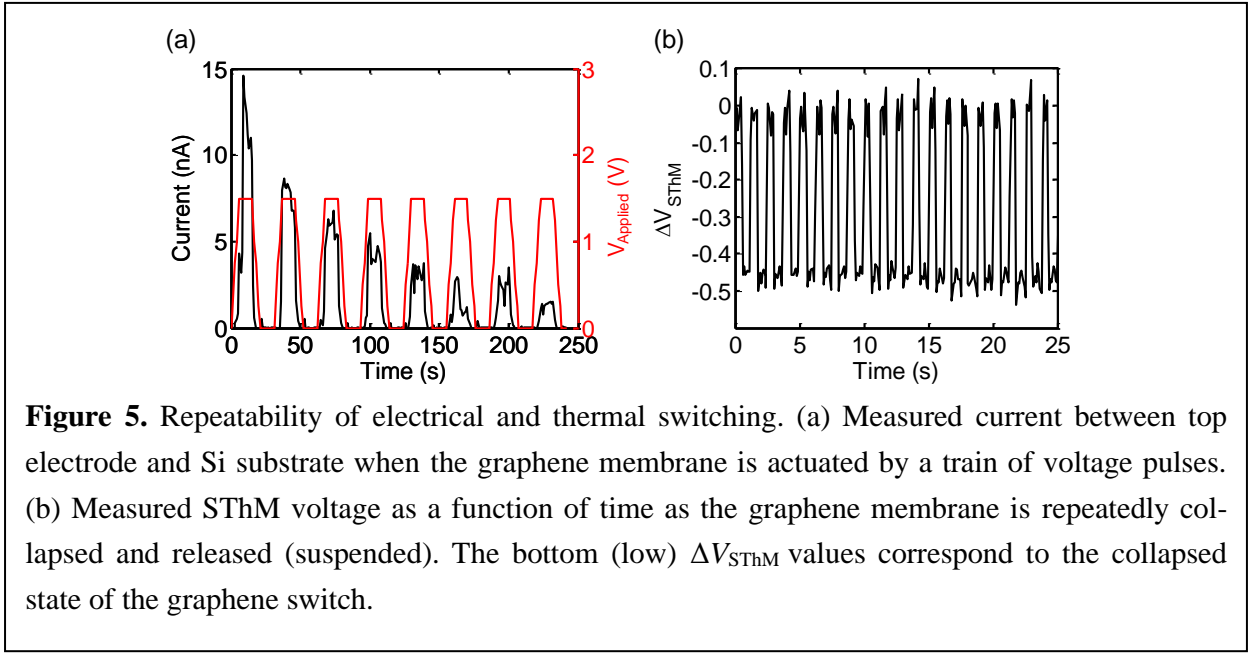

**Figure 5.** Repeatability of electrical and thermal switching. (a) Measured current between top electrode and Si substrate when the graphene membrane is actuated by a train of voltage pulses. (b) Measured SThM voltage as a function of time as the graphene membrane is repeatedly collapsed and released (suspended). The bottom (low) $\Delta V_{SThM}$ values correspond to the collapsed state of the graphene switch.

To directly measure thermal resistance switching, we used the SThM probe tip to mechanically collapse and suspend the graphene membrane, by pushing on the graphene near the electrode edge. Figure 5(b) displays this measurement, clearly demonstrating repeatable thermal cycling, where the changes in $\Delta V_{SThM}$ correspond to changes in the heat flow path between the SThM tip and the Si substrate. These measurements show a switching frequency of 0.8 Hz, which is limited by the scanning rate of the SThM tip. However, previous studies have shown NEMS resonators made with suspended graphene can reach the MHz range [32], suggesting that such thermal switches may be among the fastest achievable. Speed of switching may not be limited by mechanical resonance alone; we must also consider the thermal time constant of the device, or how quickly the device can heat and cool. The thermal time constant is given by $\tau \approx \rho c_p V \mathcal{R}$ where $\mathcal{R}$ is the thermal resistance, $c_p$ the material specific heat, $\rho$ the material density, and $V$ the volume of the body. Dollerman *et al.* [33] have measured 25 to 250 ns as the thermal time constant for 2 to 5 μm wide suspended graphene membranes.

12The thermal time constant will further increase when factoring in the additional thermal boundary resistance of graphene as it makes contact with the underlying substrate.

## 4. Conclusions

In conclusion, we designed and fabricated the first graphene-based NEMS thermal switch and demonstrated multiple, reversible electrical and thermal switching at low electrostatic actuation voltages ($< 2$ V). We have also realized the first practical demonstration of thermal metrology for a NEMS device using active mode SThM. A compact analytical model shows the thermal performance of our device can be optimized by adding flexible metal structures stacked on atomically thin membranes. This is experimentally demonstrated using graphene switches patterned with overlying serpentine chrome lines, whose change in heat flux between the off and on states is increased to 22%. Additional modeling demonstrates the effect of active thermal switching on a system under varying heat load, and illustrates the trade-off between reducing temperature swing and average temperature as a function of thermal switching ratio. The results of this work demonstrate the feasibility of implementing high thermal conductivity materials in nanoscale thermal switches and are essential for the future design and implementation of active thermal management for densely-integrated systems.

**Acknowledgments**

Fabrication and experiments were performed at the Stanford Nanofabrication Facility (SNF) and Stanford Nano Shared Facilities (SNSF), funded under National Science Foundation (NSF) award ECCS-1542152. This work was supported by the NSF Engineering Research Center for Power Optimization of Electro Thermal Systems (POETS) with cooperative agreement EEC-1449548, and by ASCENT, one of the six centers in JUMP, a Semiconductor Research Corporation (SRC) program sponsored by DARPA. MEC acknowledges support from the NSF Graduate Fellowship under grant No. DGE-1656518. SMB and EP acknowledge support from the Stanford SystemX Alliance.

**ORCID IDs**

Michelle Chen https://orcid.org/0000-0001-6508-9163

Miguel Muñoz Rojo https://orcid.org/0000-0001-9237-4584

Aditya Sood http://orcid.org/0000-0002-4319-666X

Stephanie Bohaichuk http://orcid.org/0000-0003-2705-4271

Christopher Neumann https://orcid.org/0000-0002-4705-9478

Andrew Alleyne https://orcid.org/0000-0002-1347-9669

Eric Pop https://orcid.org/0000-0003-0436-8534

# Supplementary Information

# Graphene-Based Electromechanical Thermal Switches

Michelle E. Chen[1], Miguel Muñoz Rojo[2,3], Feifei Lian[3], Justin Koeln[4,5], Aditya Sood[1], Stephanie M. Bohaichuk[3], Christopher M. Neumann[3], Sarah G. Garrow[4], Andrew G. Alleyne[4], Kenneth E. Goodson[1,6], Eric Pop[1,3]

[1] Dept. of Materials Science & Engineering, Stanford University, Stanford, CA 94305, U.S.A.

[2] Dept. of Thermal &Fluid Engineering, University of Twente, Enschede, 7500 AE, Netherlands.

[3] Dept. of Electrical Engineering, Stanford University, Stanford, CA 94305, U.S.A.

[4] Dept. of Mechanical Sci. & Eng., University of Illinois at Urbana-Champaign, Urbana, IL 61801, U.S.A.

[5] Dept. of Mechanical Engineering, University of Texas at Dallas, Richardson, TX 75080, U.S.A.

[6] Dept. of Mechanical Engineering, Stanford University, Stanford, CA 94305, U.S.A.

[*]Contact: epop@stanford.edu

## S1. Monolayer graphene growth and transfer

Monolayer graphene growth was carried out on high purity (99.9%) copper foils (JX Mining) with preferentially uniform crystal orientation (100) and large grain size (>10 μm) in an Aixtron Black Magic Pro Chemical Vapor Deposition (CVD) reactor [1]. The foil was first annealed at 1060°C for 22 minutes in forming gas (Ar/$H_2$ 500 sccm/30 sccm) to remove native surface oxide. After the anneal step, $CH_4$ was additionally flowed (10 sccm) for 15 minutes for the growth step. The self-limiting CVD process yielded monolayer graphene across the foil substrate, which was verified using Raman spectroscopy.

Poly (methyl methacrylate) (PMMA) 950 A4 was spin-coated over the graphene and copper foil, with the PMMA serving as a polymer scaffold during the wet graphene transfer [2]. The foil substrate was etched away using $FeCl_3$ and the remaining PMMA/graphene stack was rinsed in deionized (DI) water, then cleaned of ions and organic impurities in sequential baths of dilute HCl/$H_2O_2$, DI water, and dilute $NH_4OH$/$H_2O_2$. The PMMA/graphene stack was then transferred to the final substrate: 540 nm thick thermally grown $SiO_2$ on highly doped (*n*+ type, 1 to 5 mΩ·cm) silicon and dried. Finally, the PMMA was removed using acetone, leaving graphene on the oxide (Figure S1(b)).

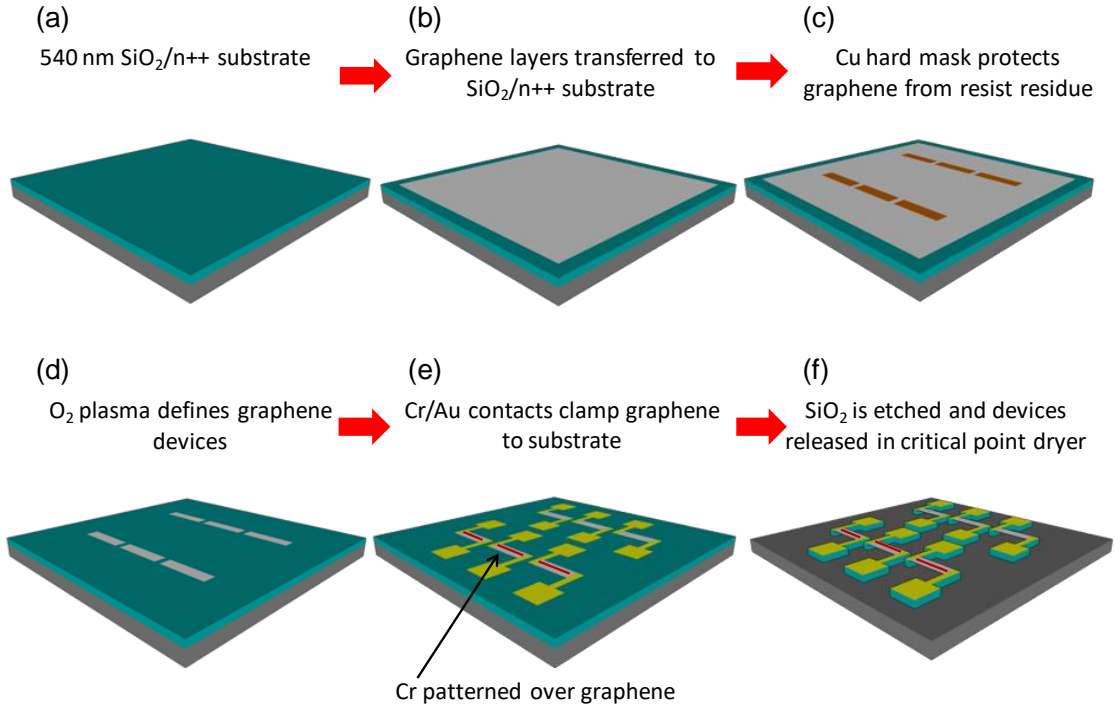

**Figure S1.** Schematic of graphene deposition, patterning, and suspension process with and without chromium patterned over the graphene.

## S2. Fabrication of suspended graphene devices

For the fabrication of suspended graphene thermal switches, two sequential polymer transfers of monolayer graphene were carried out to create an artificial bilayer stack on a 540 nm SiO$_2$/Si substrate (Figure S1(b)). The active region of the graphene devices were defined using optical photolithography, over which 40 nm of copper was deposited. This copper served as a sacrificial hard mask over the active device region (Figure S1(c)). O$_2$ plasma etching was utilized to remove the excess graphene not masked by copper. Following the O$_2$ plasma etch, the copper hard mask was removed using a commercial etchant (Transene Copper Etchant 49-1), leaving arrays of graphene microribbons free of any photoresist residue (Figure S1(d)).

For some devices, an additional photolithography and e-beam evaporation step (3 nm Cr) were used to pattern metal lines over the graphene channels. Cr/Au top electrodes and contact pads (3 nm / 40 nm) were deposited by e-beam evaporation which clamped the graphene to the substrate (Figure S1(e)). The metal electrodes immediately adjacent to the graphene were 5 or 10 μm wide and 70 μm long to minimize heat sinking to the contact pads during measurements. The deposited electrodes and contact pads served as an etch mask for the SiO$_2$. Using 20:1 buffered oxide etch, the regions of SiO$_2$ not masked by the Cr/Au regions were removed, releasing the graphene membranes (Figure S1(f)). By varying the etch duration we removed ~500 nm or 540 nm of the underlying oxide for devices that were either electrically insulating or electrically conductive cross-plane when switched. Finally the released graphene membranes were rinsed in deionized (DI) water and isopropanol (IPA) before transfer into a critical point dryer. The resulting graphene structures clamped underneath Cr/Au electrodes were thus suspended over SiO$_2$ legs without collapse (Figure S1(f)).

## S3. SThM measurement and calibration

The thermal measurements were conducted in Asylum Research MFP-3D AFM system equipped with a scanning thermal microscopy (SThM) module [3] from Anasys Instruments®. We employ the SThM approach due to its sub-100 nm spatial resolution and because the temperature measurements are performed on a metal pad, which cannot be achieved with an alternative method like Raman thermometry [4].

The V-shaped SThM tip (Figure 2) is equipped with a built-in palladium resistor on SiN (Anasys GLA), whose electrical resistance changes with temperature. We utilize active mode SThM: the tip is placed in contact with one electrode of the thermal switch and the Pd resistor electrically heated with voltage $V_{in}$ (Figure S3) to a constant temperature prior to a switching event. The constant steady state electrical resistance of the SThM tip is extracted from the voltage reading of an external, balanced Wheatstone bridge circuit voltage reading, $V_{ws,bal}$ while the device is off (Figure S3). Figure S2 shows the calibration of electrical resistance of the SThM tip as it is electrically heated. We were careful to apply the same constant force set-point during all our measurements to account for effects of thermal boundary resistance. Under thermal steady state conditions, the electrical heating of the tip is counterbalanced by atmospheric convection, radiation, and thermal boundary resistance of the tip on the device top electrode until a constant temperature is reached [3, 5]. Once this steady-state temperature condition is reached, the SThM tip undergoes no further thermoresistive change. The Wheatstone bridge was then balanced by adjusting the potentiometer within the Wheatstone bridge circuit until $V_{ws,bal}$ approached ~0 V.

The SThM tip senses the temperature change at the electrode as the graphene switches between the off (suspended) and on (collapsed) states (Figure 2(a-b)); when the graphene switch is collapsed (switched on), the increase in cross-plane heat flow causes a corresponding temperature decrease and electrical resistance change in the SThM probe tip. This correlates with a voltage reading $V_{ws}$ taken relative to the initial, balanced Wheatstone bridge circuit voltage reading the difference of which is $V_{SThM} = V_{ws} - V_{ws,bal}$ (Figure 2(c)). The final value $\Delta V_{SThM}$ is measured between the states in which the graphene is fully suspended and fully collapsed. Based on this reading, we calculate the change in electrical resistance of the SThM tip, and hence the change in temperature at the top contact of the thermal switch (Supplementary Equation 1). This measurement is proportional to the ratio of thermal resistance of the switch between the off and on state (main text Equation 2). It is important to note that even if $V_{ws,bal}$ is not perfectly balanced at 0 V, the measured thermal switching ratio of our switch is not affected. This is due to the fact that the switching ratios are proportional to the temperature change $\Delta T$ measured by SThM, which in turn is proportional to $|\Delta V_{SThM}|$; by taking the absolute difference in voltage between the device's off and on state, we negate any initial voltage imbalance within the Wheatstone bridge for our subsequent temperature change calculation.

The Wheatstone bridge equation is used to calculate the change in electrical resistance of the SThM probe given the voltage reading across the Wheatstone bridge.

$$V_{WS} = V_{in} \cdot \frac{R_2}{(R_2 + R_{pot})^2} \cdot \frac{\Delta R_{probe}}{R_1} \qquad (1)$$

$$R_{probe} = R_{probe,0} + \Delta R_{probe} \qquad (2)$$

where $R_{pot}$ is the resistance of the potentiometer, $R_1$ and $R_2$ are 1 kΩ resistors on the Wheatstone bridge, and $R_{probe,0}$ is the probe resistance under balanced Wheatstone bridge conditions.

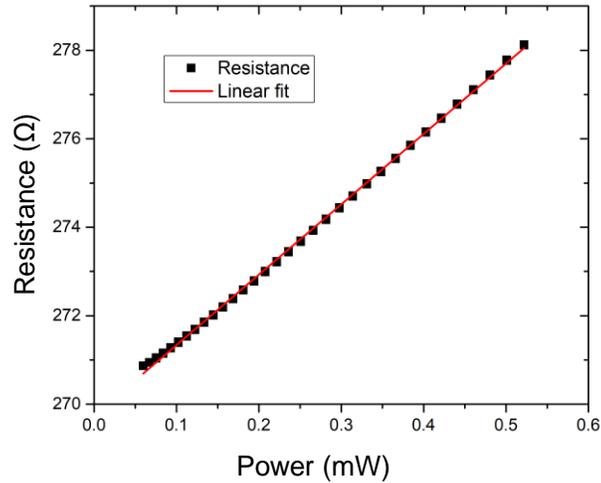

**Figure S2**. Resistance calibration of SThM tip as it is electrically heated.

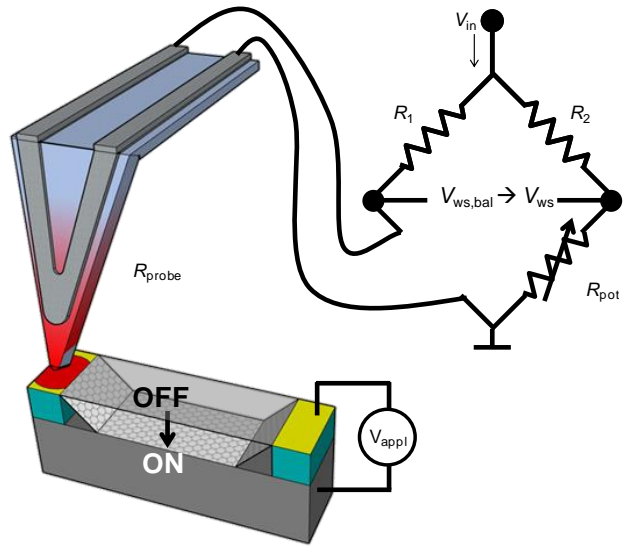

**Figure S3**. Schematic of SThM measurement set up with Wheatstone bridge circuit.

A finite element model of the graphene thermal switch device was constructed to evaluate the dependence of the measured on/off ratio on the SThM tip location. The model provides upper and lower bounds for the possible on/off ratio of the switch, which is highest when the heat source is located closest to the edge of the suspended graphene.

The simulated device dimensions are similar to experimental devices. In the off-state (suspended), the device consisted of 10 µm wide bilayer graphene suspended between two 540 nm tall SiO$_2$ pillars spaced 10 µm apart. The pillars were 15 µm long (extending 5 µm beyond the edge of the graphene width on one side) and 5 µm wide. The graphene on the pillars was coated with 40 nm thick Au. The Si substrate was approximated as 25 µm x 30 µm, with 10 µm thickness.

To simulate the on-state (collapsed) of the device, an additional SiO$_2$/Si pillar was added as support in the center of the graphene. Adding a raised pillar is easier to model than deforming the graphene down to touch the substrate, but maintains the same heat transport. The raised pillar spanned the full graphene width and consisted of 40 nm of SiO$_2$ on 500 nm of Si, with a length of 8 µm. This left 1 µm of suspended graphene adjacent to each metal contact.

The model uses the heat transfer module in COMSOL Multiphysics, which solves Fourier's law of heat conduction in steady state, with the thermal conductivities listed in Table S1. A thermal boundary resistance TBR = $2 \times 10^{-8}$ m$^2$K/W was applied at all interfaces, which is typical for graphene/metal, graphene/SiO$_2$ [6], and SiO$_2$/Si interfaces [7]. The SThM tip heat source was modelled as a Gaussian spot with peak power 375 µW and a radius of 100 nm. Heat loss due to air convection is included on the top surfaces with a coefficient of 10 W/(m$^2$·K), and an ambient temperature of 300 K.

**Table S1 – Model Parameters**

|  | $k$ [W/(m·K)] |
|---|---|
| Graphene: in-plane | 1200 |
| cross-plane | 5 |
| SiO$_2$ | 1.4 |
| Si | 140 |
| Au | 100 |

The peak temperature on the metal contact pad's surface is higher the closer the SThM tip is placed to the edge of the metal (nearest the graphene channel), as shown in Figure S4(a). Taking the ratio of the maximum temperature rise between the off (suspended membrane) and on (collapsed membrane) states, we estimate the on/off ratio in Figure S4(b). The on/off ratio is maximized by placing the tip (or a hot microelectronic component) closest to the thermal switch. The value is saturated below 200 nm, which is the approximate diameter of the SThM tip.

Given that the SThM tip was not placed exactly at the electrode edge, but was within 1 µm from it, the error in the measured on/off ratio is estimated to be <0.02.

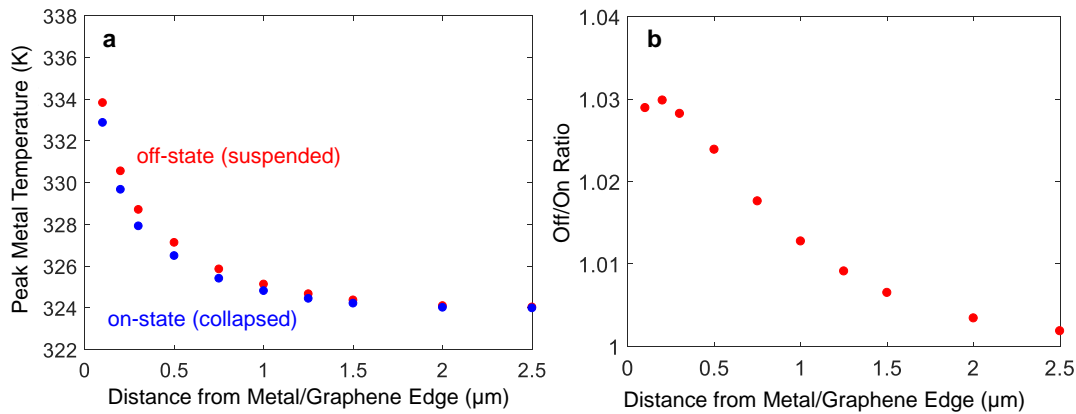

**Figure S4.** Finite element simulations of the graphene thermal switch. (a) The temperature of the metal electrode directly underneath the SThM tip, as well as the temperature difference between the on and off states, are highest if the tip is placed closer to the electrode edge. (b) The on/off ratio varies by ~0.02 with SThM tip placement on the electrode.

## S4. Electrostatic model

We calculate the pull-in voltage of the suspended graphene membrane based on the electrostatic pull-in model from Bao et al. [8]. Figure S4(a) depicts the geometric schematic of the device. The vertical deflection $h_0$ at the center of a doubly clamped suspended graphene beam clamped at $x = \pm L/2$ is calculated by using the formula

$$\frac{\varepsilon_0}{2}\left(\frac{\varepsilon_r}{\varepsilon_r d_1 + d_2}\right)^2 V^2 L^2 = 8T_0 t h_0 + \frac{64}{3}\frac{Et}{L^2(1-v^2)}h_0^3 \quad (3)$$

where $\varepsilon_0$ is the permittivity of free space, $\varepsilon_r \sim 3.9$ is the relative dielectric constant of $SiO_2$, $d_1 \sim 500$ nm is the trench depth, $d_2 \sim 0$ nm is the thickness of $SiO_2$ at the bottom of the trench, $V$ is the applied voltage, $L \sim 15$ μm is the length of the suspended beam, $T_0 = [E/(1-v^2)](\Delta L/L)]$ is the stress in the membrane at equilibrium, $\Delta L \sim 0$ μm is the relative elongation of the suspended beam due to tension or slack, $t \sim 0.68$ nm is the thickness of 2 layers of graphene [8], $v \sim 0.165$ is the Poisson ratio of graphite in the basal plane [8], and $E \sim 1$ TPa is the Young's modulus of graphene [8]. The off state of the device is defined by the condition where vertical membrane deflection $h_0 < d_1$ the trench depth (Figure S5(a)). We define the pull-in condition, or the point the device is switched on, as the state where $h_0 = d_1$. That is, the pull-in voltage $V_{PI}$ is the value of $V$ which yields the solution $h_0 = d_1$ for Eq. 3 [Figure S5(b)].

At $V > V_{PI}$, we approximate the length of graphene which is in contact with the bottom of the trench as $L_c$. $L_c$ is a parabola chord approximated from a parabolic profile of a membrane with deflection $h_0$ (Figure S5(c)). The parabolic profile given by Bao et al. is [8]:

$$y = 4h_0\left(\frac{x}{L_{\text{trench}}}\right)^2 \quad (4)$$

For the pull-in condition where $h_0 > d_1$, $L_c$ is the chord corresponding to the solution of Eq. 4 setting $y = d_1$. Using the boundary conditions of $(L/2, h_0)$ and $(L_c/2, h_0 - d_1)$ to solve Eq. 3, we approximate the length of graphene in contact with the substrate as:

$$L_c = L \cdot \sqrt{1 - \frac{d_1}{h_0}} \quad (5)$$

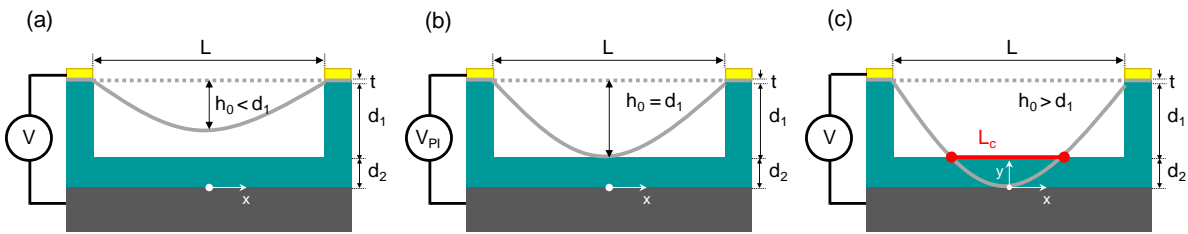

**Figure S5** Cross-section schematic of doubly clamped membrane with actuation voltage (a) below membrane pull-in voltage, $V < V_{PI}$; (b) at membrane pull-in voltage, $V = V_{PI}$; (c) above pull-in voltage, $V > V_{PI}$.

## S5. Effect of adding Cr lines and their geometry

We integrate metal patterns on top of the graphene (Figure 3) to create a flexible metal-graphene switching membrane with enhanced thermal conductivity while maintaining low actuation volt-age. This approach utilizes the high Young's modulus and intrinsic strength of the underlying graphene [8] to mechanically support the thin metal shape (3 nm Cr). This thickness of chromium approaches the fundamental limits of solid matter [9] which would otherwise not have the mechanical integrity for free-standing suspension. We patterned 3 nm of Cr over the graphene beam in different geometries and compared the average switching ratio of our 27 switches consisting only of 2 stacked graphene layers. Chromium is selected as the metal layer due to its good adhesion to graphene, which is necessary for withstanding subsequent processing.

Two patterns are investigated: 2 parallel straight lines, and a serpentine line, as shown inset to Figure S5. The thermal switching ratio of both is improved compared to the graphene-only switch, with the parallel line exhibiting a switching ratio of 1.10, and serpentine line performing with the best switching ratio of 1.22 (Figure S6).

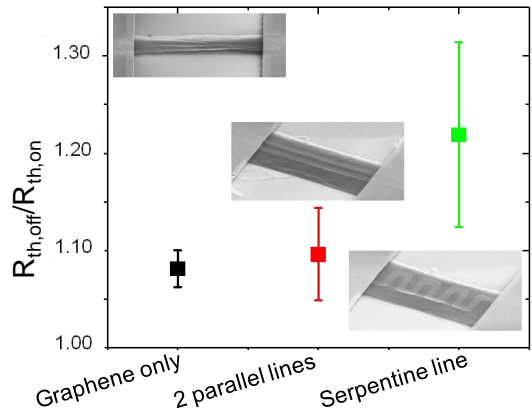

**Figure S6**. Average thermal switching ratio of graphene-only thermal switches, and graphene patterned with 3 nm of chromium in geometries of two parallel lines and a serpentine zig-zag line. Error bars indicate the standard deviation from the average switching ratio.

## S6. System level modeling

System level modeling of the benefits of a general thermal switch was carried out for heating densities of 100 W/cm$^2$ and 50 W/cm$^2$ [Figures 4(c) and S7(a), respectively]. Figure S7(b) lists the parameter values considered for a switch with a micron thick silicon heat sink (substrate). The on-state thermal resistance of the thermal switch is set as constant, and the thermal switching ratio is varied in reference to the on-state (i.e. the off-state thermal resistance is varied). If the switching ratio is high, then the temperature swing amplitude decreases, highlighting the main benefit of thermal switching over passive heat sinks. However, Figures 4(c) and S7 show that as the thermal switching ratio increases, there is also significant and undesirable temperature rise as a result of a higher thermal resistance. Thus, more critical than the thermal switching ratio, is ensuring a low overall on-state thermal resistance of the thermal switch in order to maintain practical operating temperatures of electronic devices and systems, thus avoiding temperature-induced failure. This is in contrast to design considerations for an electrical switch, where minimal off-state electrical leakage is desirable. For thermal switches, minimizing the overall thermal resistance in both the off- and on-state is critical and must be optimized by geometry or materials selection.

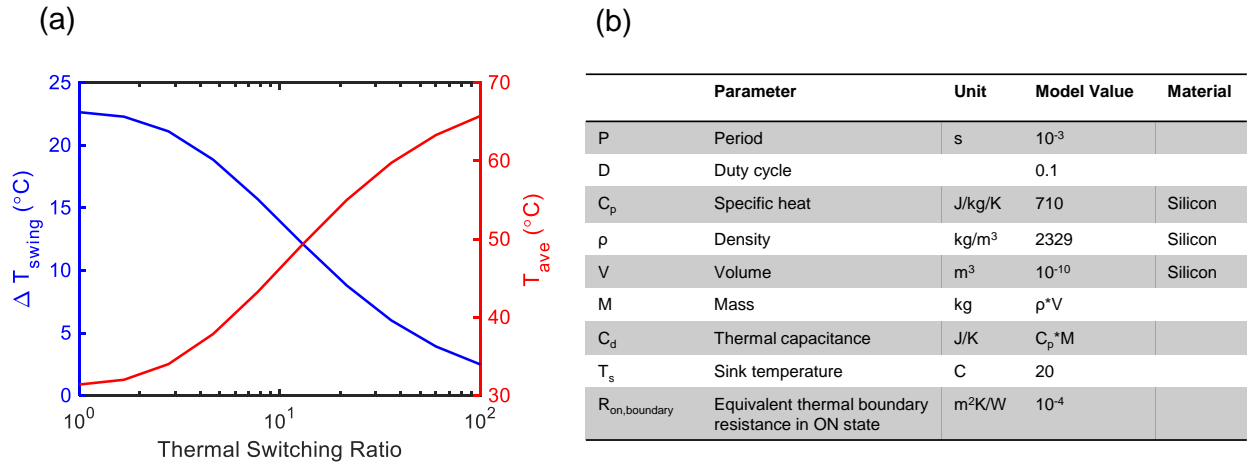

**Figure S7**. (a) Calculated temperature swing amplitude and average temperature vs. thermal switching ratio for a 1 × 1 cm system with 50 W heat operating at kHz frequency for (b) the listed parameters. The system considered corresponds to the thermal circuit shown in Figure 4 of the main text.

## Supplementary References: